\documentclass[aps,prl,reprint,groupedaddress]{revtex4-1}
\usepackage{graphicx}
\usepackage{amsmath}
\usepackage{bm}
\usepackage{hyperref}

\begin{document}

\title{Coulomb interactions and the spatial coherence of femtosecond nanometric electron pulses}

\author{Stefan Meier}
\email[]{stefan.m.meier@fau.de}
\author{Peter Hommelhoff}

\affiliation{Department of Physics, Friedrich-Alexander-Universität Erlangen-Nürnberg (FAU), D-91058 Erlangen, Germany, EU}

\date{\today}
\begin{abstract}
The transverse coherence of electrons is of utmost importance in high resolution electron microscopes, point-projection microscopes, low-energy electron microscopy and various other applications. Pulsed versions of many of these have recently been realized, mostly relying on femtosecond laser-triggering electron emission from a sharp needle source. We here observe electron interference fringes and measure how the interference visibility becomes reduced as we increase the electron bunch charge. Due to the extremely strong spatio-temporal confinement of the electrons generated here, we observe the visibility reduction already at average electron bunch charges of less than 1 electron per pulse, owing to the stochastic nature of the emission process. We can fully and quantitatively explain the loss of coherence based on model simulations. Via the van Cittert-Zernike theorem we can connect the visibility reduction to an increase of the effective source size. We conclude by discussing emittance, brightness and quantum degeneracy, which have direct ramifications to many setups and devices relying on pulsed coherent electrons.
\end{abstract}

\maketitle

\begin{figure}
	\centering
	\includegraphics[width=1\linewidth]{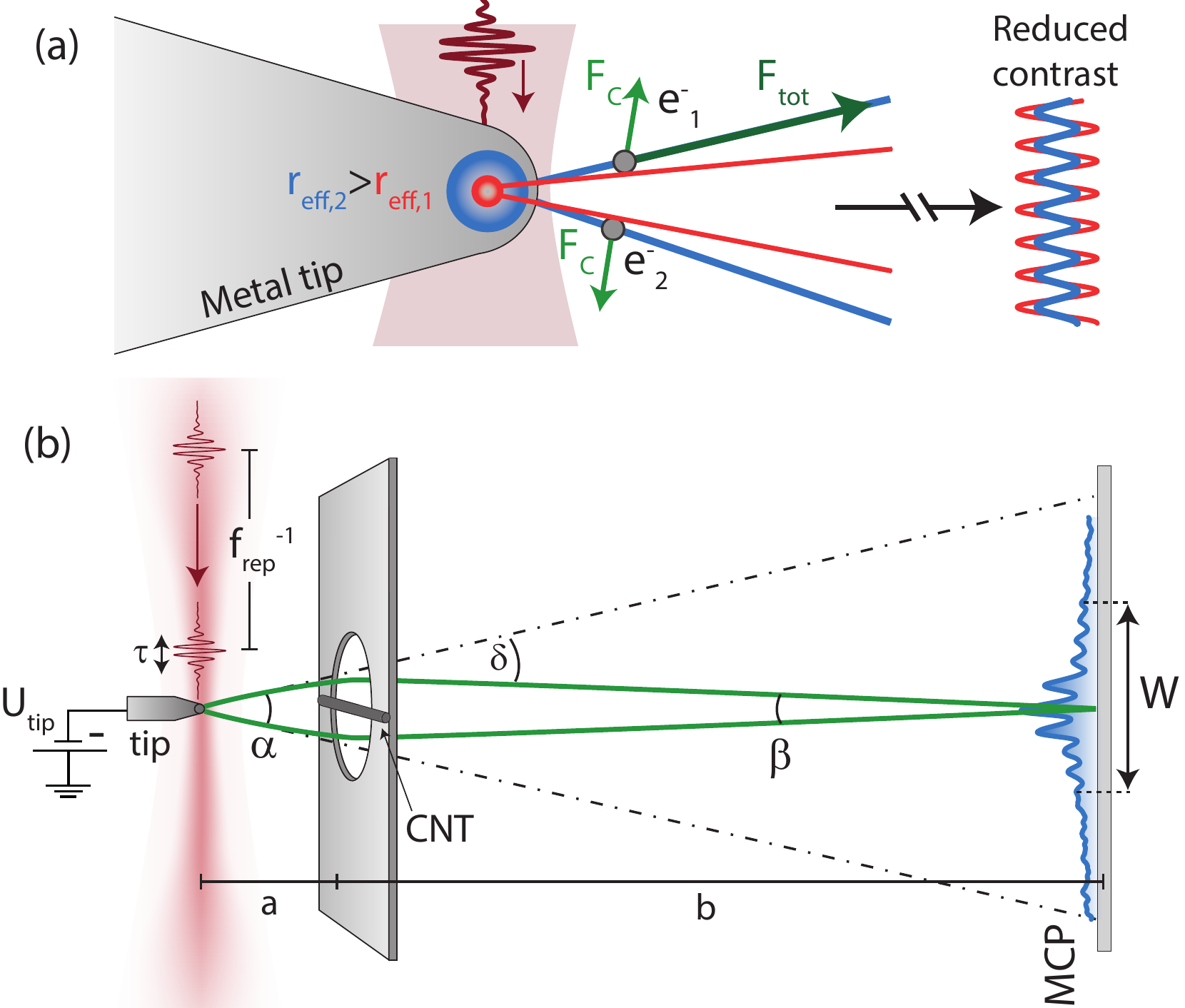}
	\caption{(a) Coulomb interactions in few-electron pulses lead to a reduced contrast in electron interference experiments. This can be described by an increase of the effective source size: the red area inside the tip is effectively spread out to the blue area due to the repulsive forces (green arrows). The blue lines represent electron trajectories where the electrons repel each other by Coulomb forces \(F_C\). Together with the acceleration due to the static tip voltage, the total force \(F_\text{tot}\) acts on the electron. The red lines show the respective trajectories without Coulomb interaction. (b) Scheme of the experimental setup. Electrons are triggered by laser pulses with duration \(\tau=6\,\)fs and repetition rate \(f_\text{rep}=80\,\)MHz from a tungsten needle tip, biased with \(U_\text{tip}\). A single-walled carbon nanotube (CNT) is approached to the tip to a distance \(a\) (roughly \(1\,\mu\)m) and acts as an electrostatic biprism. Electron wave interference fringes are detected on an MCP screen at a distance \(b=6.7\,\)cm from the CNT. The green lines depict two outgoing parts of an electron wave under a mutual angle \(\alpha\) that are deflected by an angle \(\delta\) as they pass the CNT and overlap at the screen, forming the interference pattern of width \(W\). \label{fig:setup}}
\end{figure}

A key factor for a well-controlled electron beam is its coherence, meaning the self-similarity of the electrons contained in the beam. Coherence is closely associated with the point-like nature of the source, while both properties are in turn associated with the focusability of the beam. These closely related properties of modern electron sources, coherence and nanometer source size, are used in electron lithography or atomically resolving electron microscopes, for example, as they enable the smallest focus of an electron beam \cite{Spence2013}. These properties are equally important in research, including more imaging-oriented topics such as electron holography \cite{Lichte2008,Longchamp2013} or in basic research such as the observation of the fermionic equivalent to the Hanbury-Brown Twiss experiment \cite{Kiesel2002,Kuwahara2021}.\\
To generate coherent electron beams in practice, besides aberration-free electron optics \cite{OKeefe2008} highly coherent electron emitters are needed \cite{Spence2013}. Experiments showed that electrons emitted from sharp tungsten needle tips \cite{Cho2004}, carbon nanotubes \cite{Schmid1997}, as well as from single-atom tips \cite{0957-4484-20-11-115401} show high spatial coherence.\\
If the emitted current becomes larger and larger, Coulomb interactions between the electrons start to distort the beam. This becomes significantly important when the emission is also highly confined in time, namely by triggering the electron emission with ultrashort laser pulses \cite{Hommelhoff2006_2,Hommelhoff2006,Ropers2007,Arbouet2018}.\\
In previous work, we have shown that also ultrashort electron pulses from tungsten needle tips show high spatial coherence, but this was in a regime with well below one electron per pulse \cite{Meier2018}. Here, we systematically investigate the transition to the multi-electron per pulse regime. Although Coulomb interactions in electron beams have long been studied, their effect on laser-triggered electron pulses is subject of intense current research. It was shown that space charge limits the brightness of picosecond pulses containing several hundreds to thousands of electrons \cite{Kuwahara2016}. Another experiment demonstrated an increased spot size in the interaction region of an ultrafast transmission electron microscope (UTEM) due to Coulomb interactions \cite{Bach2019}. In work focusing on strong field emission, it could recently be shown that multi-electron effects strongly affect the resulting electron spectra during the emission process \cite{Schoetz2021}. Great progress has also been made on Coulomb interactions in electron emission in theory and simulations: Resolution limits for ultrafast electron diffraction \cite{Ischenko2019} or aberrations caused by space charge in a TEM \cite{Denham2021} are only a few recent examples for simulations dealing with high bunch charges from a few femtocoulomb up to a few picocoulomb. For the highly relevant regime of a few electrons per pulse, single-trajectory simulations were carried out, showing how Coulomb interactions can limit the brightness of cold field emission \cite{Kruit2010} and pulsed electron emission \cite{Cook2016}. However, there are no experiments on the behavior of the spatial coherence of ultrafast electron pulses from needle sources containing more than one electron.\\
In this work, we investigate the contrast of interference fringes of ultrashort pulsed electron emission as a function of the emitted pulse charge, shown schematically in Fig. \ref{fig:setup} (a). Electrons laser-emitted from a needle tip pass a beam splitter and interfere on a screen [Fig. \ref{fig:setup} (b)]. The contrast or visibility of the observed fringe pattern equals the degree of spatial coherence, given by the van Cittert-Zernike theorem \cite{Lichte2008,Ehberger2015,Meier2018}. Our experiments show a decreasing visibility with increasing emitted current, which is a direct effect of the Coulomb interaction between electrons being closely spaced near the tip surface shortly after the emission as we will show. The effect already occurs at an average of below one electron per pulse, since two-electron events already show up due to Poisson statistics. The loss of visibility and thus spatial coherence can be interpreted as a growth of the effective size of the electron emission source [Fig. \ref{fig:setup} (a)].\\
In the experiment, electrons are emitted by focusing laser pulses from a Ti:Sapphire oscillator on a sharp tungsten needle tip with an apex radius of 5 to 10 nm. The laser pulses are spectrally centered around \(780\,\)nm and have a pulse duration of \(\tau=6\)\,fs at a repetition rate of \(80\,\)MHz. The laser beam is focused down to a focus size of \(1.1\,\mu\)m using an off-axis parabolic mirror with a focal length of 15 mm. The experiment takes place inside of an ultra-high vacuum chamber with a pressure \(<1\cdot10^{-9}\,\)hPa. The tungsten tip is biased with a voltage between \(-30\) to \(-60\)\,V to accelerate the electrons away from the tip after emission.\\
A single-walled carbon nanotube (CNT) spanning across a hole on a TEM-grid is approached closely to the tip; typically the distance \(a\) is a few \(\mu\)m down to around \(1\,\mu\)m [Fig. \ref{fig:setup} (b)]. The CNT acts as an electrostatic biprism for the emitted electrons \cite{Cho2004,Ehberger2015,Meier2018}. After passing the CNT, the split electron waves interfere and this interference pattern is recorded with a micro channel plate (MCP) plus phosphor screen placed at a distance \(b=67\,\)mm. The electrons passing the CNT are deflected by an angle \(\delta\), which depends on the distance \(a\) between tip and CNT as the potential bends around the CNT \cite{Weierstall1999}. By choosing a distance \(a\), the mutual angle \(\alpha\) between two partial rays that overlap at the center of the resulting interference pattern can be selected. The observable width \(W\) of the interference pattern depends on \(\alpha\) and \(b\) in the case of a negligibly small biprism radius \cite{0022-3735-14-6-001}.\\
In (electron) optics, the van Cittert-Zernike theorem states that for a spatially extended incoherent source, the angular coherence function \(\mu^\text{SC}(\alpha)\) is given by the Fourier-transform of the source intensity distribution \(I(x)\) \cite{vanCittert1934,Zernike1938,Lichte2008,Saleh2019}:
\begin{align}
\mu^\text{SC}(\alpha)=\int_{-\infty}^{\infty}\exp\left(i\frac{2\pi}{\lambda}\cdot\alpha x\right)I(x)\mathrm{d}x.
\label{eq:vcz}
\end{align}
\(\mu^\text{SC}\) is also called the \textit{degree of spatial coherence} and equals the interference visibility if two parts of the beam, with an angular distance of \(\alpha\), interfere. Here, \(\lambda\) is the de Broglie wavelength of the electrons and \(x\) the spatial coordinate in the source plane. Based on this we determine the size of the source intensity distribution by observing the resulting interference pattern. The source intensity distribution is typically called \textit{the effective source size} \(r_\text{eff}\) in electron optics. In an earlier experiment we measured \(r_\text{eff}=0.65\,\)nm for multiphoton-photoemitted electrons from tungsten needle tips with physical tip radii of \(\sim 7\,\)nm in the same setup \cite{Meier2018}.\\
\begin{figure*}
	\includegraphics[width=0.9\linewidth]{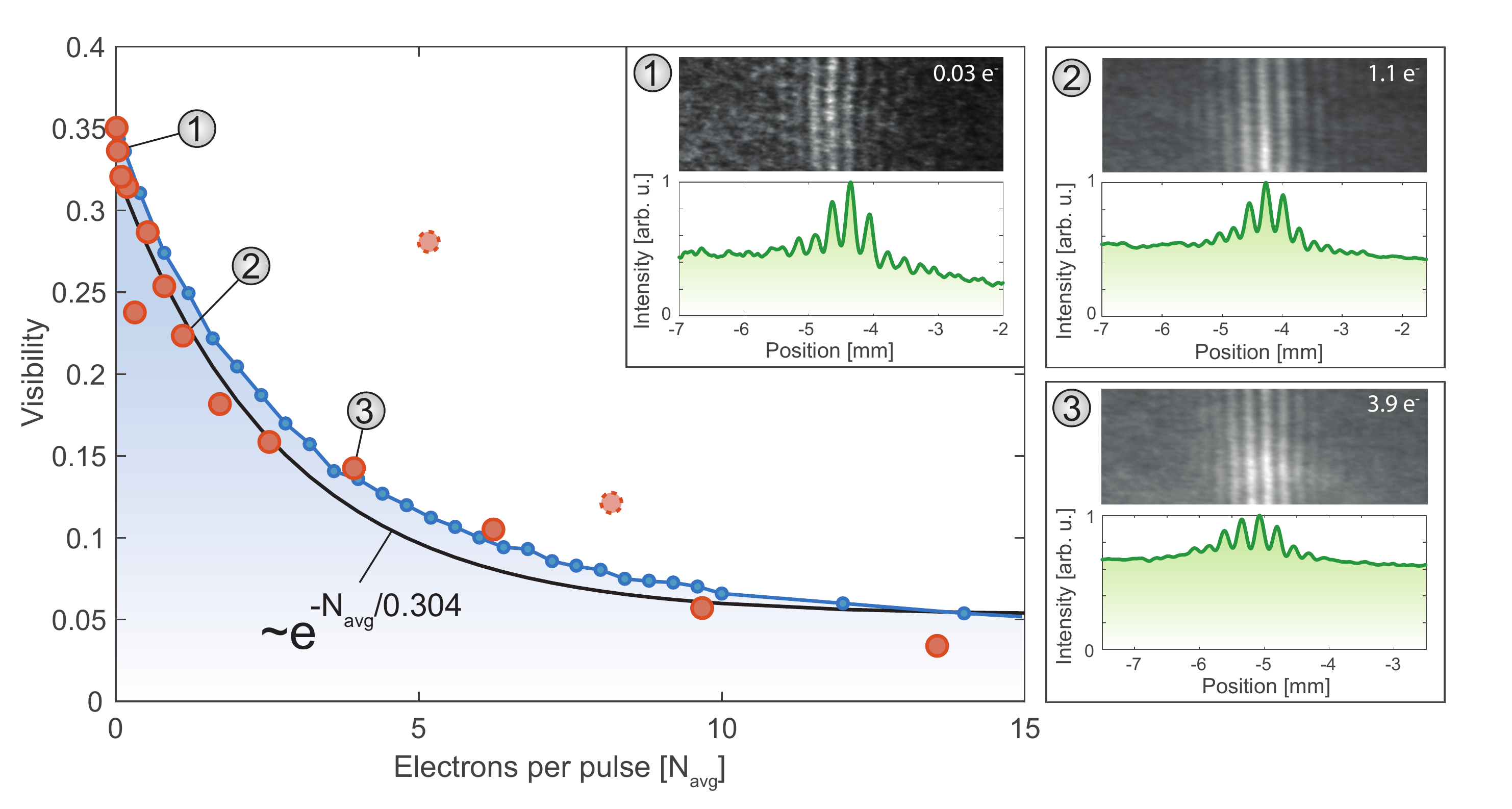}%
	\caption{The experimentally observed electron matter wave interference visibility decreases with increasing current from the tip (orange data points). The blue curve represents the calculated visibility based on the effective source size increased due to Coulomb repulsion, matching the experimental data very well. The black line represents an exponential fit. Insets 1 to 3 show example electron detector images of the measurement (top) and vertically integrated line profiles (green), from which the visibility was evaluated. In 1, an example for low emission current (\(\sim 0.03\) electrons per pulse) is shown; the visibility of the central fringe is 0.34. Insets 2 and 3 show measurements at higher currents of \(\sim 1.1\) and \(\sim 3.9\) electrons per pulse. The respective line profiles show visibilities of 0.22 and 0.14.\label{fig:figExpData}}
\end{figure*}
In the present experiment, we record interference patterns for increasing laser powers for one specific angle \(\alpha\), set by a fixed distance \(a\). The emitted current from the tip is recorded for each step and follows a power law scaling \(I\propto P_L^n\) due to the multiphoton-photoemission process, where \(P_L\) is the laser power. Here, \(n\) denotes the multiphoton order and is in the range of 2-3.\\
We recorded 16 such interferograms and evaluated the visibility for a range of emitted currents, corresponding to 0.01 to 13.6 electrons per pulse, see Fig. \ref{fig:figExpData}. Clearly, our measurements show a reduction of visibility from 0.35 for the smallest current down to 0.03 for the largest current. The insets 1-3 in Fig. \ref{fig:figExpData} show results for emitted currents of 0.38 pA, 14 pA and 50 pA, i.e. 0.03, 1.1 and 3.9 electrons per pulse. We calculate the visibility as \(\mathcal{V}=\frac{I_\text{max}-I_\text{min}}{I_\text{max}+I_\text{min}}\), as explained in more detail in \cite{supplementary_information}.\\
We now show that this reduction of visibility with increasing pulse charge is a direct manifestation of Coulomb interactions of the emitted electrons, strongly confined in space and time. The electrons repel each other directly after emission, leading to an effectively increased source, as we see from the van Cittert-Zernike theorem: With Eq. \eqref{eq:vcz} applied to a Gaussian intensity distribution \(I(r)\), the degree of spatial coherence reads \cite{Lichte2008}:\\
\begin{align}
\mu^\text{SC}=\exp\left[-\left(\frac{\pi\alpha r_\text{eff}}{\lambda_\text{dB}}\right)^2\right].
\label{eq:degree_of_spatial_coherence}
\end{align}
Hence, the visibility depends on the mutual angle \(\alpha\), the electron de Broglie wavelength \(\lambda_\text{dB}\) and the effective source size \(r_\text{eff}\). The tip-to-sample distance as well as the bias voltage \(U_\text{tip}\) remained unchanged in the experiment, so \(\alpha\) and \(\lambda_\text{dB}\) are fixed. Hence, the only variable that may vary due to Coulomb interaction of the electrons is the effective source size \(r_\text{eff}\), a direct consequence of the reduction in contrast.\\
We carried out a numerical simulation to further confirm this notion. We numerically integrate the equation of motion for electrons, treated as point particles and emitted within one pulse \cite{supplementary_information}. Here, we vary the number of electrons and average over many initial conditions. From the results, we can extract an effective source size, based on the emitted current. Using Eq. \eqref{eq:degree_of_spatial_coherence}, we calculate the visibility for the simulated effective source sizes. Here, the mutual angle \(\alpha\) is expressed by \(\alpha\approx W/b\). For the experiment shown in Fig. \ref{fig:figExpData}, we measured \(W=3.3\,\)mm at the screen. The simulated visibilities are shown in blue in Fig. \ref{fig:figExpData}. In the experiment, the contrast is additionally reduced due to finite spatial resolution and vibrations between tip and CNT. Therefore we multiplied the blue data with a factor of 0.4 to match with the experimental data for the lowest currents, as the influence of Coulomb repulsion is minimal there. We find excellent agreement of the simulated visibility values to the experiment.\\
The visibility decays nearly exponentially with increasing current in both experimental data and simulations. As a sanity check we turn off the Coulomb interaction in the simulation and find that the effective source size stays constant at \(\sim 0.6\)\,nm for all currents, resulting in a constant visibility.\\
This quantitative agreement between simulation and experiment clearly identifies Coulomb interactions between emitted electrons as the main cause for a decreasing spatial coherence with increasing number of emitted electrons per pulse.\\
\begin{figure}
	\includegraphics[width=\linewidth]{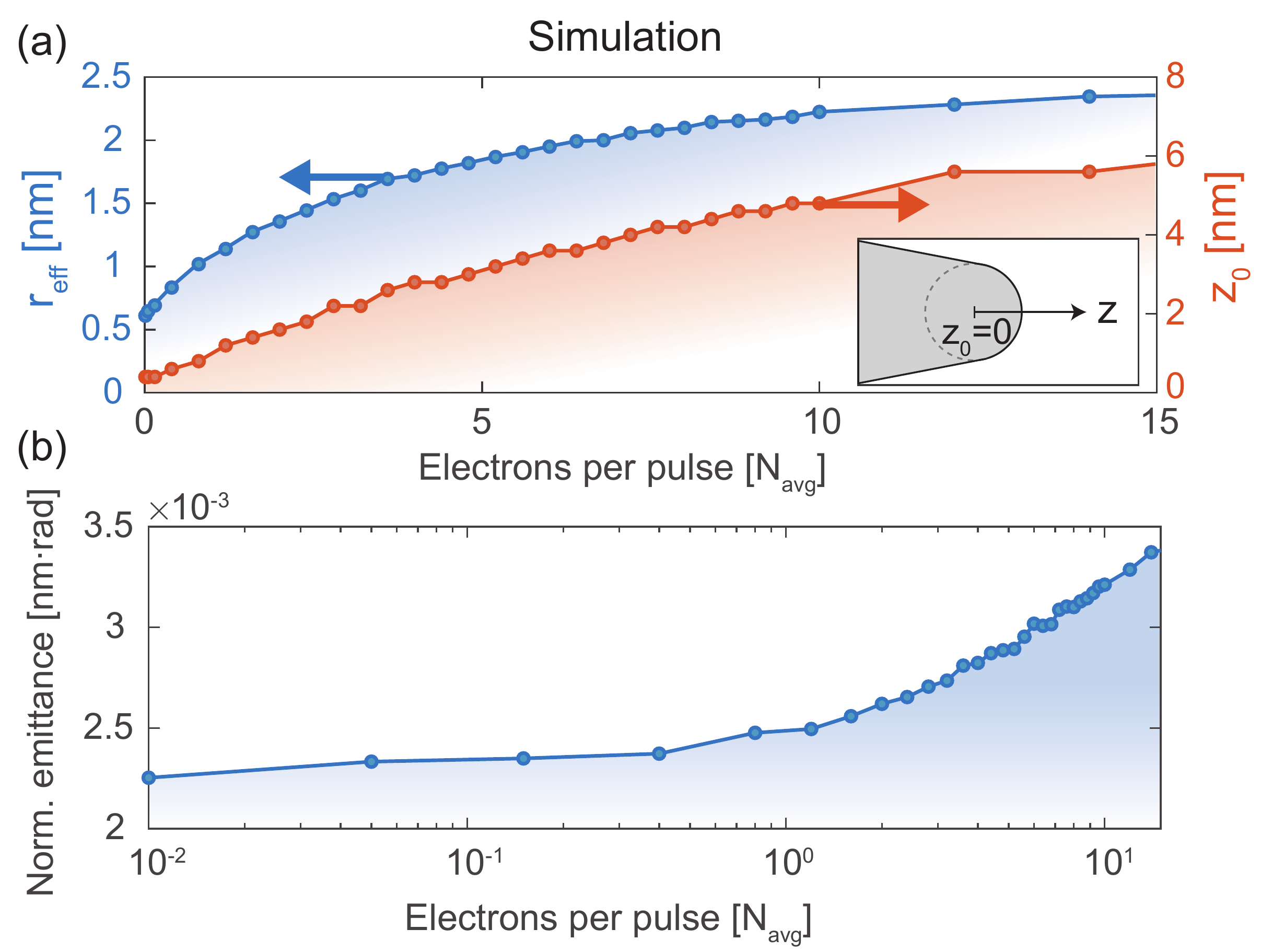}%
	\caption{(a) Simulated effective source size as function of the emitted current, shown in blue. The Coulomb repulsion affects the electron trajectories such that the effective source position (orange curve) is shifted towards the positive z-direction, see inset. We expect this effect to lead to current-dependent system settings in a TEM, for example. (b) Simulated growth of the normalized rms-emittance. \label{fig:FigSim}}
\end{figure}
The simulated source size as function of the emitted current is shown as the blue curve in Fig. \ref{fig:FigSim} (a). For low currents the simulation yields \(r_\text{eff}\sim 0.6\)\,nm, which increases up to \(\sim 2.3\)\,nm for 14 electrons per pulse on average. In addition we obtain that the virtual source plane shifts towards the outside of the tip (orange curve) \cite{supplementary_information}. Fig. \ref{fig:FigSim} (b) shows the emittance of the simulated electron beam. The emittance represents the defining parameter considering the spatial properties of an electron beam. We find that Coulomb interactions are also directly reflected in an increase of the emittance, also noticeably starting around one electron per pulse. We note that this dependence underscores existing numerical work \cite{Cook2016}. At best in any linear optics system, the emittance is preserved during beam propagation, which we show here is not the case anymore as soon as Coulomb interactions are taken into account.\\
In contrast to bosonic particles like photons, there is a fundamental limit for the maximum achievable brightness of an electron beam due to the Pauli exclusion principle \cite{Gabor1961}. Because of this, the degeneracy of an electron beam given by the actual brightness over the Pauli limited brightness \(\delta=B/B_\text{max,Pauli}\) can not exceed unity \cite{Silverman1987a}. From our simulations, we find that values as large as \(\delta>1\cdot10^{-2}\) close to the tip become feasible for ultrashort electron pulses even with Coulomb interactions, exceeding the highest values reported for DC sources of \(\delta\sim1\cdot10^{-4}\) \cite{Kiesel2002}. These insights bode extremely well for measuring quantum limits of electron sources \cite{Lougovski2011}.\\
The simulations show that a smaller emission area leads to a smaller emittance, despite Coulomb interactions. However, smaller tips do not need to be favorable in terms of brightness, as also an increasing energy spread must be taken into account. As the temporal spread of the pulse cannot be easily reversed, a smaller tip can therefore even lead to a smaller far-field brightness after some propagation of the beam. The ideal source parameters are hence highly system dependent.\\
We note that the observed loss in spatial coherence cannot be attributed to either \textit{quantum decoherence} or classical \textit{dephasing} \cite{Schlosshauer2019}. The latter effect could, in principle, be reversed by some additional measurement, which is not possible in our case. The former is understood as a coupling of the quantum mechanical system to a decohering environment \cite{Hornberger2003,Sonnentag2007,Kerker2020}, again different to what we have. Here, instead, the electrons repel each other in a random manner, different from shot to shot, and thus accumulate a variable interaction phase. This variable phase reduces the contrast of the beam after incoherent integration. This implies that if one could post-select each event according to the number of electrons emitted, we could obtain further insights on the coherence of each N-electron event class, an intriguing future experiment.\\
To summarize, we have shown experimentally and quantitatively that the spatial coherence of laser-triggered electrons decreases as a function of electrons emitted per pulse. The reduction of spatial coherence can be interpreted as a current-dependent increase of the effective source size of the emitter. For the extreme spatio-temporal confinement generated in this work, an average value of a mere 1.4 electrons per pulse suffices to let the visibility drop to 65\% of the value unaffected by interactions, which already corresponds to a doubling of the effective source size. We could show that for these parameters the driving force for the decrease in coherence are Coulomb interactions between the emitted electrons close to the tip. We stress that our results are very general as they are independent of tip material, laser wavelength etc.. They need to be considered in any application or measurement with a strong spatiotemporal confinement of electrons, e.g. for time-resolved electron microscopy \cite{Arbouet2018}. Clearly, the number of electrons per pulse before Coulomb repulsion sets in notably can be largely increased with increased tip radius and increased laser pulse duration. Last, our work sets the stage to proceed to measuring fermionic quantum effects in the interaction of electrons highly confined in space and time.
\begin{acknowledgments}
The authors would like to thank Dr. Manuel Nutz and Prof.~Alexander Högele for the production of the CNT-samples and Prof. Klaus Hornberger for fruitful discussions. This research was supported by the European Research Council (Consolidator Grant NearFieldAtto and Advanced Grant AccelOnChip) and the Deutsche Forschungsgemeinschaft (SFB 953 and TRR 306).
\end{acknowledgments}

\end{document}